# Fluorescence and FRET based mercury (II) sensor

Dr. Jaba Saha, Sudip Suklabaidya, Dr. Jayasree Nath, Dr. Arpan Datta Roy, Bapi Dey,

Dr. Dibyendu Dey, Prof. D. Bhattacharjee, Dr. Syed Arshad Hussain*

Department of Physics, Tripura University, Suryamaninagar – 799022, Tripura, India

* Corresponding author

Email: sa_h153@hotmail.com, sahhussain@tripurauniv.in

Ph: +917005694182 (M), +91381 2375317 (O)

Fax: +913812374802

**ABSTRACT**

A mercury (II) sensor has been proposed based on Fluorescence Resonance Energy Transfer (FRET) between N,N'-dioctadecyl thiacyanine perchlorate (NK) and Octadecyl rhodamine B chloride (RhB). Out of these two molecules NK is sensitive to $Hg^{2+}$ ions due to presence of sulfur atom in it. Accordingly, presence of $Hg^{2+}$ ions affects the NK fluorescence as well as FRET from NK to RhB. Our results showed that NK fluorescence intensity and FRET efficiency linearly decreases with increase in $Hg^{2+}$ ion concentration. With proper optimization present system under investigations can be used to sense $Hg^{2+}$ ions in aqueous solution with detection limit of 9.13 ppb. Advantage of this present system is that it is very simple compared to the other FRET based system and also it works under aqueous environment. This method has also been tested using real lake water and satisfactory results were obtained.

## 1. Introduction

Heavy metal pollution is a serious concern in the present era as it has posed a serious threat to public health across the globe. This has made rapid detection of detrimental heavy metal ions in water as one of the most important issues because of their significant role in chemical, biological and environmental assays [1-10]. Mercury (II), one of the heavy metal ions, has been recognized as the most perilous, toxic, highly carcinogenic and hazardous pollutants; and its rampant contamination arises from a variety of natural and anthropogenic sources [11-14]. It is capable of causing serious health complications because of its close affinity to the thiol groups of enzymes and proteins. These complications may be listed as motion and central nervous system disorders, brain damage, immune system dysfunction, cardiovascular diseases, cognitive and motion disorders. It may also result in the impairment of visual and hearing senses, blindness and death for regular intake of $Hg^{2+}$ even at a low concentration [15-20]. Mercury and its compounds have been ranked third by The United States Comprehensive Environmental Response Compensation and Liability Act (CERCLA) in the "priority list of hazardous substances" [21, 22]. Solvated mercury, which is one of the most invariable inorganic forms, is, highly toxic due to its easy water solubility [23, 24]. $Hg^{2+}$ in the contaminated natural water with its deleterious effect on human health offers the scientific community a challenging task of developing rapid, cost-effective, environmentally friendly, sensitive and selective $Hg^{2+}$ detection methods to detect this harmful pollutant in aqueous environment. Therefore, the topic regarding detection of $Hg^{2+}$ in aqueous environments has assumed a serious significance in the scientific mind so that its resultant health complications could be avoided [25-27].

Researchers throughout the world have shown immense interest to design $Hg^{2+}$ sensors based on various principles. Literature survey revealed that most commonly reported methods are such as Charge Transfer (CT), Electron Transfer (ET), Energy Transfer (eT), Förster

Resonance Energy Transfer (FRET), Through bond energy Transfer (TBET), Cold Vapor Atomic Absorption Spectrometry (CVAAS), Inductively Coupled Plasma Mass Spectrometry (ICP-MS), Ultraviolet and visible spectrophotometry (UV–vis) and electrochemical methods [28-48]. Further details about various mercury detection systems have been discussed in the review paper by P. D. Selid et. al. [49].

However, fluorescence spectroscopic methods is considered as the most powerful tool for $Hg^{2+}$ as well as other heavy metal ion detection because of its sensitivity, simplicity, inexpensive and on-site detection proficiency [41-48]. Most of the fluorescence sensors are based on either fluorescence enhancement (turn-on) or fluorescence quenching (turn off) in presence of $Hg^{2+}$ [13-15]. These are mainly a single fluorescence signal based sensors, where, the signal may be readily disturbed by environmental perturbations e.g. pH, temperature, solvent, excitation source fluctuation, sensor concentration etc [25-27]. This can be overcome by employing ratiometric FRET based fluorescence sensing system, where, two fluorescence signals are used. Intensity of one fluorescence signal will increase and at the same time intensity of the other will decrease [1-6]. Ratio of these two signals will give idea about the sensing efficiency. As a result effect of different environmental perturbations has been eliminated automatically [50-53].

There exits few reports where, FRET phenomenon has been employed to design $Hg^{2+}$ sensors [1-6]. However, most of these FRET based sensors work in organic/mixed solvents. Accordingly they cannot be used in most of the environmental as well as biological applications [54-56]. Therefore, it is very important to design FRET based $Hg^{2+}$ sensing system which can work under aqueous environment [57]. Nanoparticle [36, 37] as well as quantum dot [29, 30, 35] based $Hg^{2+}$ sensing system employing FRET are suffering from toxicity, selectivity, inadequate resolution in aqueous environment. For most of the FRET based systems special fluorescence probes have to be designed which require rigorous and complex synthesis process [18, 53].

In the present work we report a simple FRET and fluorescence based $Hg^{2+}$ sensing system, where two dyes NK (N,N'-dioctadecyl thiacyanine perchlorate) and RhB (Octadecyl Rhodamine B chloride) were used as FRET pair. Here, we have incorporated both the molecules onto thin film using Langmuir-Blodgett (LB) as well as Spin Coating (SC) techniques at desired mixing ratio. After that the film is dipped onto the $Hg^{2+}$ solution (sample solution) followed by its spectroscopic analysis. Our system was capable of detecting $Hg^{2+}$ with detection limit of 9.13 and 11.7 ppb in LB and SC film respectively.

Main advantage of the present system is that the process is very simple and no complex synthesis process is required. Also the system work under aqueous environment.

## 2. Experimental

### *2.1. Material*

NK was purchased from Hayashibara Biochemical Laboratories Inc. RhB and mercury chloride ($HgCl_2$) was obtained from Sigma Chemical Co. USA. All the salts $ZnCl_2$, $FeCl_3$, $AlCl_3$, $MgCl_2$, $CaCl_2$, and $CuCl_2$ were obtained from Thermo Fisher Scientific India Pvt. Ltd. All materials were used without further purification. The dyes as well as mercury ions ($Hg^{2+}$) used in our studies are positively charged. Chloroform (purity >99%, Sigma Chemical Co. USA), and ultrapure water (resistivity 18.2 MΩ) were used as solvents.

### *2.2. SC film preparation*

The SC films were prepared by using a commercial SC Unit (SCU 2005A, APEX Instrument Co, India) operating at motor speed fixed at 6000 rpm for 60 s. In this technique (SC), chloroform solutions of dyes are placed on a spinning solid support, followed by rotation at desired rpm.

*2.3. Isotherm study and LB film preparation*

LB instrument (APEX-2000C, India) was used for study of monolayer characteristics and LB film deposition. Either pure water or aqueous solution of $HgCl_2$ was used as the subphase as per requirement. Chloroform solution of NK, RhB or NK-RhB mixture was spread onto the subphase for isotherm measurement as well as for LB film deposition onto solid substrate. The concentrations of dyes used in this study were $0.5\times10^{-3}$ M. LB film deposition was done at 10 mN/m surface pressure. All the experiments were done at ambient condition. Details about LB film technique have been given elsewhere [58].

*2.4. Brewster Angle Microscopic (BAM)*

If someone interested to investigate ultrathin films on transparent dielectric substances Brewster Angle Microscopy is an ideal tool for him. When a p-polarized light incident at $53.1^{\circ}$ (Brewster angle) on air-water interface, principally no reflection going to occur. So, in absence of any organic layer i.e., only in presence of water ‘black background’ will be observed. However, if a very thin layer of organic substances is present, refractive index will change. Consequently, reflection will occur. This reflected light may be utilized to obtain information about floating film. BAM instrument (Accurion nanofilm EP4-BAM, Serial No. 1601EP4030) was used for the present study. BAM images were recorded for pure NK, RhB as well as their mixture in air water interface in addition to $Hg^{2+}$ subphase. After spreading the sample solution onto LB trough and waiting for sometimes to evaporate the solvent, images were recorded consistently at various surface pressures.

*2.5. Spectroscopic Study*

Spectroscopic measurements (absorption and fluorescence) were carried out by Absorption Spectrophotometer (Lambda-25, Perkin Elmer) and Fluorescence Spectrophotometer (LS-55, Perkin Elmer). All the Fluorescence spectra were recorded with excitation wavelength 430 nm.

## 3. Results and discussions

*3.1. FRET between NK and RhB*

Thiacyanine dye NK and xanthene dye RhB used in the present study are highly fluorescent with high quantum yield [59]. Sufficient spectral overlap exits between the fluorescence spectra of NK and absorption spectra of RhB (Figs. S1 and S2). Therefore, in principle these two dyes are suitable for FRET to occur. NK will act as donor and RhB will act as an acceptor. We have already investigated FRET phenomenon using these two dyes [59]. It has been observed that FRET occurred between these two dyes under certain conditions [59]. In homogeneous solution there was no FRET [59]. However, FRET occurred when these molecules were assembled onto restricted geometry such as clay surface as well as onto thin films [59].

Here, we have selected these two fluorophores as FRET pair to design FRET based ratiometric $Hg^{2+}$ sensor. Pearson's Soft Hard Acid Base (SHAB) principle states that hard acids prefer to combine with hard bases and soft acids prefer to combine with soft bases resulting stable complex formation [60]. $Hg^{2+}$ ions belong to Pearson's soft Lewis acid. On the other hand, there exist two sulfur atoms in NK molecule, which belongs to soft base category. Therefore, it is expected that $Hg^{2+}$ will form complex with NK through the interaction with its sulfur atom. This complex formation should affect the energy transfer process between NK and RhB. This is because the change in orientation as well donor-acceptor distance largely affects the extent of FRET [61-65]. Also to best of our knowledge ratiometric $Hg^{2+}$ sensor employing these two dyes has never been reported.

Based on our previous investigations we have planned to study FRET between NK and RhB assembled onto thin film in order to design FRET based $Hg^{2+}$ sensor [61,66]. For ultrathin film formation LB and SC techniques have been employed.

Pure NK, RhB, and NK-RhB (50:50 molar ratio) mixture fluorescence spectra assembled onto LB films and SC films are shown in supporting information (Fig. S3). All the measurements have been performed with excitation wavelength $\lambda_{ex}$=430 nm, which is very near to the absorption maxima of NK [59, 67]. At this wavelength, NK absorption was maximum and RhB absorption was almost negligible. This is because almost no absorption i. e direct excitation of RhB molecules occurred at this excitation wavelength (Fig. S3). However, interestingly the fluorescence spectrum of NK-RhB mixture for both the films, NK fluorescence decreases and RhB fluorescence increases in comparison to their pure counterpart. This indicates that energy transfer occurred from NK to RhB. Related fluorescence spectra are given in Fig. S3 and the spectroscopic features are listed in Table-1.

From recorded excitation spectra (inset of Fig. S3) it is affirmed that RhB fluorescence is predominantly because of the light absorption by NK and subsequent transfer it to RhB. The FRET efficiency was found to be 53% (for SC film) and 49% (for LB film) which was calculated using the following equation [59,68]

$$E = 1 - \frac{F_{DA}}{F_D};$$

Here, $F_{DA}$ = fluorescence intensity of the donor in the presence of acceptor

$F_D$ = fluorescence intensity of the donor in the absence of the acceptor.

*3.2. Monolayer characteristics at air water interface*

In an effort to design FRET based $Hg^{2+}$ sensor, two dyes NK and RhB has been selected as FRET pair and planned to investigate the FRET phenomenon involving these two dyes as well as $Hg^{2+}$ assembled onto thin films. LB and SC techniques have been used for ultrathin film formation. According to Pearson's SHAB principle it is expected that $Hg^{2+}$ will interact with NK molecules [26]. In order to have idea about the interaction between $Hg^{2+}$ and dye molecules, here we have studied the pressure−area (π−A) isotherms of the monolayer of NK, RhB and NK-RhB mixture in absence (Fig. 1a) and presence (Fig. 1b) of $Hg^{2+}$ using LB technique.

The isotherm of pure NK on water subphase started with an initial lift off area of 1.50 $nm^2$ and rises steeply until collapse pressure is reached at 59.9 mN/m. In case of pure RhB on water subphase the initial lift of area is 1.62 $nm^2$ and surface pressure rises up to 39.3 mN/m upon compression. As a whole, the shapes and the nature of π-A isotherms of NK and RhB are in good agreement with the previous reports [59,69,70]. However, NK–RhB isotherm has been shifted to larger area per molecule region with respect to both the pure isotherms. For the mixed isotherm the initial lift off area was 1.97 $nm^2$. This enlargement in area per molecule for NK–RhB mixed isotherm is a clear indication about the strong interaction between the constituent molecules [71]. Due to positively charged nature of NK as well as RhB, there exits electrostatic repulsion between NK and RhB in the mixed monolayer film.

To explore the effect of $Hg^{2+}$ on the floating monolayer, we spread NK, RhB, or NK-RhB mixture solution onto the water subphase containing $HgCl_2$ ($10^{-6}$ M) and recorded the corresponding isotherms. For all the case waiting time after spreading was 30 minute to allow interaction between spread molecule and $Hg^{2+}$ in the subphase. Interestingly, NK and NK-RhB mixture isotherm in presence of $Hg^{2+}$ showed marked change with respect to the isotherm recorded in pure water subphase. Here, for both the isotherms area per molecules is shifted to the larger area with initial lift off area 2.01 and 2.26 $nm^2$ respectively. Area per molecule observed at 15 mN/m surface pressure was 1.23 and 1.34 $nm^2$ respectively for NK and NK-RhB mixed isotherm. On the other hand, RhB isotherm in presence of $Hg^{2+}$ showed almost negligible change with respect to the same measurement in water subphase. Observed change in NK and NK-RhB isotherm in presence of $Hg^{2+}$ indicated that there exists strong interaction between $Hg^{2+}$ in the subphase and the molecule (NK) onto the floating layer. As mentioned earlier, sulfur atom present in the NK molecule is a soft base and $Hg^{2+}$ is a soft acid. Accordingly, they will interact strongly as per SHAB rule [47]. Due to this interaction in presence of $Hg^{2+}$, NK-$Hg^{2+}$ complex were formed at air water interface. The size of the complex is larger in comparison to that of pure NK. Accordingly, the area per molecule shifted to the larger area in case of NK and NK-RhB mixed isotherm. At the same time, for RhB, there exists almost no interaction between RhB and $Hg^{2+}$. Accordingly, there was no scope for $Hg^{2+}$ to come onto air water interface. Consequently, area per molecule did not affect much in presence of $Hg^{2+}$. Our later investigations of BAM images of floating monolayer also support this.

It has also been observed that incorporation of $Hg^{2+}$ onto the floating monolayer (inset of Fig. 1b) largely affected the compressibility of the floating film [72]. In presence of $Hg^{2+}$ compressibility of NK-RhB mixed film decreases significantly. This may be due to the fact that incorporation of $Hg^{2+}$ makes the floating film more rigid. Earlier reports also suggested that incorporation of metal ions enhances the rigidity of floating films [73, 74]. Therefore, lowering of compressibility of the NK-RhB mixed film prepared onto $Hg^{2+}$ subphase also indicated the incorporation of $Hg^{2+}$ onto the floating NK-RhB mixed films.

To have visual evidence about the incorporation of $Hg^{2+}$ onto LB film we have investigated the NK, RhB and NK-RhB floating film in absence and presence of $Hg^{2+}$ using BAM. BAM is an authentic tool generally used to get information regarding the morphological changes of the floating monolayer [66]. Accordingly, visual evidence of complex formation or interaction in the multi component LB film can be obtained through BAM studies [75].

Fig. 2 shows the BAM images recorded for NK, RhB and NK-RhB mixture LB film in presence and absence of $Hg^{2+}$. All the images presented here were taken at 10 mN/m surface pressure. BAM images of the same recorded at other different surface pressures are shown in Fig. S4.

From the BAM images taken at empty water surface (fig. not shown here) fully dark images has been observed. This is because; light cannot be reflected without any floating material [66]. BAM image of $Hg^{2+}$ aqueous subphase without any dye spread onto the subphase likewise exhibited black image (fig. not shown here). Due to the water solvated nature of $Hg^{2+}$ ions, they remained within the subphase and almost no $Hg^{2+}$ molecules appear onto the air water interface [23, 31].

BAM images of NK and RhB on pure water subphase showed almost similar morphology (Fig. 2a and b). At the beginning at lower surface pressure the domains were smaller. Although, they were visible throughout the image and almost uniformly distributed (Figs. S4a and S4b). This corresponds to the gaseous phase of the isotherm at the beginning of the isotherm and just after the beginning of compression. With increase in surface pressure upon compression the molecules at air water interface come close to each other resulting formation of larger domains and almost compact film at the interface (Fig. 2a and b). This indicates the phase change in the floating film at higher surface pressure [70]. In case of NK-RhB mixture the surface morphology of the floating film was different (Fig. 2c). Here the domain organization was different even at the beginning of compression i.e. very low pressure (Fig. S4c). With increase in surface pressure almost no further change was observed (Fig. 2c). Here, small islands of domains were observed throughout the image. Also, lots of blank surfaces were observed. This is because NK as well as RhB molecules were cationic and there exist repulsive interaction between these two molecules in the mixed film. Our previous isotherm studies also supported the repulsive interaction between them. Due to this repulsive interaction both NK and RhB moves away from each other. Several NK molecular domains come close to each other. Same thing happened for RhB also. As a result, colony of NK and RhB molecular domains formed throughout the film. However, it was not possible to distinguish between NK and RhB domain colony since both NK and RhB domains were almost identical (Fig. 2a and b). Also the resolution of BAM instrument did not allow this.

BAM images recorded onto $HgCl_2$ subphase showed marked changes compared to that measured in pure water subphase (Fig. 2). In case of NK monolayer BAM images showed large domains with leaf like structures having different brightness across the domains (Fig. 2d). Even at very low surface pressure such structures were observed (Fig. S4d). As the surface pressure increases, the number of domains in addition to surface coverage increases (Fig. 2d). It has already been mentioned that according to SHAB principle $Hg^{2+}$ interacted strongly with the floating NK and form NK-$Hg^{2+}$ complex at air water interface [26]. Due to this interaction, domain morphology changes at the interface in presence of $HgCl_2$ surface at the subphase [72].

At the same time, BAM images of RhB monolayer recorded in presence of $HgCl_2$ almost identical to the same recorded in absence of $Hg^{2+}$ (Fig. 2e). This may be due to the fact that there exists almost no interaction between the floating RhB molecule and $Hg^{2+}$ in the subphase (Fig. S4e).

Again in case of NK-RhB mixed monolayer, presence of $HgCl_2$ in the subphase showed marked change on the BAM images. Here, chain/wire like domains were observed throughout the image (Fig. 2f) at all the surface pressures (Fig. S4f). However at higher surface pressure the domains become more compact and several domains connect to each other. Here also the observed change may be due to the interaction of $Hg^{2+}$ at subphase and NK in the mixed film. As a result $Hg^{2+}$ come onto the subphase and affects the domain morphology.

Therefore, it can be concluded that BAM investigation exhibits convincing visual proof that $Hg^{2+}$ successfully incorporated onto the NK as well as NK-RhB mixed film.

*3.3. Fluorescence spectroscopic study of NK in thin film*

According to Pearson's SHAB rule it was expected that sulfur atom present in NK molecule will interact with the $Hg^{2+}$ [60]. This might affect the fluorescence property of NK. Accordingly, we have prepared NK monolayer LB film and dipped the same into $HgCl_2$ aqueous solution of different concentration ranging from $6\times10^{-8}$ M to $4.5\times10^{-7}$ M. After that the fluorescence spectra were recorded.

Fig. 3a showed the corresponding fluorescence spectra. Plot of fluorescence intensity (484 nm) as a function of $Hg^{2+}$ concentrations are also plotted in the inset of Fig. 3a. Interestingly, a decrease in fluorescence intensity of NK has been found in presence of $Hg^{2+}$. Due to strong attraction of $Hg^{2+}$ towards sulfur atom in NK, the lone pair on sulfur atom did not take part in resonance and hence the fluorescence intensity decreases [26]. It has been found that as the concentration of $Hg^{2+}$ increases, fluorescence intensity of NK decreases almost linearly (inset of Fig. 3a) [76-79].

We have studied the same with SC films also. In order to do this we have prepared spin coated film of NK onto quartz substrate. The resulting film is then dipped into $HgCl_2$ aqueous solution of different concentration ranging from $6\times10^{-8}$ M to $4.5\times10^{-7}$ M followed by their fluorescence measurement. Interestingly, here also almost similar trend was observed (Fig. 3b). The fluorescence intensity almost decreases linearly with the $Hg^{2+}$ concentrations (inset of Fig. 3b) [14].

Therefore, our fluorescence study revealed that fluorescence of NK assembled onto both LB film as well as onto SC films are highly affected by the existence of $Hg^{2+}$ in aqueous environment. Both the cases the fluorescence intensity decreased almost linearly (inset of Fig. 3a and b) with the increase in concentration of $Hg^{2+}$ in aqueous environment. This suggests that with proper calibration it's possible to design fluorescence based $Hg^{2+}$ sensor using the NK LB film as well as NK SC films.

Nowadays, fluorescent sensing has turned into a robust tool for detection of different chemicals along with biological materials [41-48]. These sensors employ a single signal and records the variation in fluorescence intensity for detection, which could be easily disturbed by the environmental and instrumental perturbations [13-15]. On the other hand, FRET process involves the ratio of two fluorescence intensities at two different wavelengths simultaneously. Here, fluorescence intensity of one signal decreases and at the same time intensity decreases for the other. Idea about sensing efficiency can be obtained from the ratio [1-6]. In such manner FRET process can be a fascinating technique to design ratiometric sensors with high selectivity and simplicity [29,30, 61-65].

FRET using NK and RhB as FRET pair has already been identified [59]. Therefore, in an attempt to design FRET based ratiometric sensor for $Hg^{2+}$ we have checked FRET process between NK and RhB in presence of various concentration of $Hg^{2+}$.

### *3.4. Effect of $Hg^{2+}$ on FRET between NK and RhB in SC and LB films*

In section 3.1 of this manuscript it has been shown that FRET has taken place between NK and RhB with energy transfer efficiency 53% and 49% for SC and LB film respectively. Also it has been observed that $Hg^{2+}$ can be successfully incorporated into LB film as well as SC films. Here to explore the effect of $Hg^{2+}$ on FRET, we have prepared NK, RhB and NK-RhB mixed (50:50 molar ratio) SC and LB films. After that the films were dipped onto aqueous solution of $Hg^{2+}$ at different concentrations followed by fluorescence spectroscopic measurement.

Fig. 4a and b show the plot of energy transfer efficiency vs. $Hg^{2+}$ concentrations for LB film and SC film respectively. Related fluorescence spectra were shown in Fig. S5a and S5b. FRET parameters for both the systems were calculated using Forster theory reported elsewhere and are given in Table 2 and 3 [64]. Interestingly, it has been found that both the SC and LB films, energy transfer efficiency decreases almost linearly with increasing $Hg^{2+}$ concentration. This gives the idea that with proper calibration it's possible to design FRET based $Hg^{2+}$ sensor using NK and RhB as FRET pair.

*3.5. Discussion*

Studies have shown that human health can be seriously affected due to contamination of highly toxic $Hg^{2+}$ in drinking water [15-20]. In general mercury can exists in three different forms: elemental mercury ($Hg^0$), ionic mercury ($Hg^{2+}$), and organic mercury complexes [80]. Mercury can transform among these three forms based on environmental conditions [49]. Therefore, all these three forms of mercury are harmful to human health. It is important for human beings to keep away from $Hg^{2+}$ contaminated water either to drink or everyday uses. For that reason, there is an extreme need to recognize $Hg^{2+}$ contaminated water [1-6].

Herein, both the fluorescence intensity of NK as well as energy transfer efficiency between the fluorophores NK and RhB are found to be dependent on $Hg^{2+}$ ion concentration in aqueous environment. Approximate linear decrease in fluorescence intensity or energy transfer efficiency was found with increasing $Hg^{2+}$ ion concentration in aqueous environment ranging from $6\times10^{-8}$ M to $4.5\times10^{-7}$ M which is equivalent to 16.26 to 121.95 ppb. Therefore, it can be suggested that with proper optimization it is possible to sense $Hg^{2+}$ in aqueous environment by observing the fluorescence intensity of NK or FRET between NK and RhB. It is already known that FRET process become a powerful tool to design various types of sensors in recent years [62,66].

Idea about the sensitivity of the proposed sensor can be obtained by calculating the Limit of Detection (LOD). It can be determined from the calibration of fluorescence as well as FRET method based on the formula LOD = 3σ/S, where, 'σ' is the standard deviation of y intercept of the regression line and 'S' is the slope of the calibration curve [24,81,82]. In the present case curve shown in Fig. 3 and 4 have been used to calculate the LOD values. The calculated values were listed in Table-4. It has been found that calculated LOD values were lower for FRET based sensing system compared to the fluorescence based sensing system. In case of fluorescence based system standard deviation of 'y intercept of the regression line' was large which resulted larger values of LOD [35, 52].

*3.6. Selectivity of the sensor*

Practically, an effective sensor should react to a specific analyte at once; otherwise the detection procedure will have error [25]. Therefore, selectivity is a significant parameter to assess the applicability of the sensor [3-4]. In this report, $Hg^{2+}$ levels in water can be identified by observing the decrease in fluorescence intensity or energy transfer efficiency. So, it is extremely expected that as the concentration of $Hg^{2+}$ changes, the energy transfer efficiency or fluorescence intensity will change with excellent selectivity.

Therefore, we examined the effect of a variety of environmentally relevant different salts on fluorescence intensity of NK as well as FRET efficiency between NK and RhB. Interestingly negligible change in fluorescence intensity (Fig. S6) and FRET efficiency (Fig. 5) occurred in presence of these salts with respect to the presence of $Hg^{2+}$. Values of energy transfer efficiency are given in Table 5. It has also been found that FRET based sensing system is better with respect to detection limit and sensitivity. It is interesting to mention that ratiometric fluorescence based sensors are beneficial than single fluorescence sensors because they employ the ratio of two fluorescence signals simultaneously in such a way that any fluctuations in intensity because of instrumental and environmental perturbations are cancelled out [62-65].

Fluorescence spectra's corresponding to selectivity study was shown in Fig. S6. All these observations indicate that our method described herein exhibits high selectivity and sensitivity towards $Hg^{2+}$ in spite of the existence all possible interfering cations. This situation may be explained according to SHAB principle. According to this principle, hard acids prefer to combine with hard bases, and soft acids prefer to combine with soft bases [60]. As $Hg^{2+}$ is a soft acid and sulfur present in NK is a soft base, the excellent selectivity towards $Hg^{2+}$ arise due to a stronger interaction between $Hg^{2+}$ and sulfur present in NK. Due to this attraction of $Hg^{2+}$ towards sulfur in NK, the lone pair on S is not available for resonance and hence the fluorescence intensity of NK decreases.

On the other hand, almost no remarkable decrease in fluorescent intensity was found for other cations. Here $Cu^{2+}$, $Zn^{2+}$, are borderline acids and $Fe^{3+}$, $Al^{3+}$, $Ca^{2+}$, $Mg^{2+}$ are hard acids while sulfur is a soft base. Therefore these cations do not have any affinity to come in contact with sulfur present in NK. Therefore, there is no obvious change in the fluorescence intensity of NK in presence of these cations and also FRET remains almost unaffected.

*3.7. Analysis of* $Hg^{2+}$ *in real water Samples.*

To demonstrate the feasibility and applicability of the planned sensing system, we have checked the proposed FRET based sensing system by using real water samples [83]. Accordingly, we have tested the proposed FRET based sensing system using natural lake water collected from three different sources. Before testing, the natural lake water was kept overnight in a beaker following filtering to remove the insoluble substances. However, no significant change in FRET efficiency was observed. This may due to the absence of $Hg^{2+}$ ions in the collected water samples. As $Hg^{2+}$ were not found in the above real water samples, the recovery experiments were used to determine the accuracy of our method. In these experiments three different amounts of $Hg^{2+}$ ions were added with the collected lake water samples. After that our proposed method was applied to detect $Hg^{2+}$ concentrations. All the experiments were repeated three times and each result shown here is an average of three sets of independent experiments. Table -6 listed the results of recovery experiments from which it is clear that samples showed excellent recovery with (RSD) values within the range of 0.6% to 2.2% and the $Hg^{2+}$ recovered was within the range 95.29%-98.80% [62, 81]

## 4. Conclusion

In summary, a simple FRET and fluorescence based $Hg^{2+}$ sensing system has been demonstrated by using LB as well as SC techniques. Our system was capable of detecting $Hg^{2+}$ with detection limit of 9.13 ppb. This method exhibits high selectivity and high sensitivity towards $Hg^{2+}$. Furthermore, the system has been fruitfully used for detection of $Hg^{2+}$ ions in real water with standard deviation within the range of 0.6-2.2%. Main advantage of this system is that the technique is very simple without any complex synthesis process as required for other FRET based system. Also the proposed system works under aqueous environment and can be used in most of the environmental and biological applications. Although, most of the reported FRET based $Hg^{2+}$ sensing system work in organic/mixed solvents.

**Acknowledgments**

The authors are grateful to DST, Govt. of India, for financial support to carry out this research work through DST project Ref: EMR/2014/ 000234 and FIST DST project Ref. SR/FST/PSI-191/2014. The authors are also grateful to UGC, Govt. of India for financial support to carry out this research work through financial assistance under UGC – SAP program 2016. The author Jaba Saha is grateful to DST-WOS-A project- Govt. of India (Ref No.: SR/WOS-A/PM-78/2016) for financial support to carry out this research work.

**Figure caption**

**Fig. 1.** Pressure–area isotherm at the air–water interface (a) without $Hg^{2+}$, 1 = 100% RhB, 2 = 100% NK, 3 = (RhB (50%) + NK (50%)), (b) with $Hg^{2+}$, 1 = 100% RhB, 2 = 100% NK, 3 = (RhB (50%) + NK (50%)). Inset of Fig. 1(b) shows the Compressibility ($C_s$) as a function of surface pressure for RhB (50%) + NK (50%)) without $Hg^{2+}$ (i) and with $Hg^{2+}$ (ii) obtained from Pressure–area isotherms.

**Fig. 2.** BAM images of NK (a) RhB (b) NK-RhB (c) in water subphase and BAM images of NK (d) RhB (e) NK-RhB (f) in $HgCl_2$ subphase. Concentrations of NK and $HgCl_2$ are fixed at $0.5\times10^{-3}$ M and $10^{-6}$ respectively.

**Fig. 3.** Variation of fluorescence intensity of NK at different concentrations of $Hg^{2+}$ in **(a)** LB films and (**b)** spin coating films**.** Concentrations of NK is fixed at $0.5\times10^{-3}$ M. Concentrations of $Hg^{2+}$ is varied from $6\times10^{-8}$ M to $4.5\times10^{-7}$ M. Excitation wavelength is 430 nm.

**Fig. 4.** Variation of FRET efficiency between NK and RhB at different concentrations of $Hg^{2+}$ in **(a)** LB films and **(b)** Spin coating films. Concentrations of NK and RhB are fixed at $0.5\times10^{-3}$ M. Concentrations of $Hg^{2+}$ is varied from $6\times10^{-8}$ M to $4.5\times10^{-7}$ M. Excitation wavelength is 430 nm.

**Fig. 5.** FRET efficiency between NK and RhB in presence of different salts (black stacks) and in presence of different salts + $Hg^{2+}$ (red stacks). Concentrations of NK and RhB are fixed at $0.5\times10^{-3}$ M. Concentration of both $Hg^{2+}$ and different salts are $4.5\times10^{-7}$ M. Excitation wavelength is 430 nm.

**Figures**

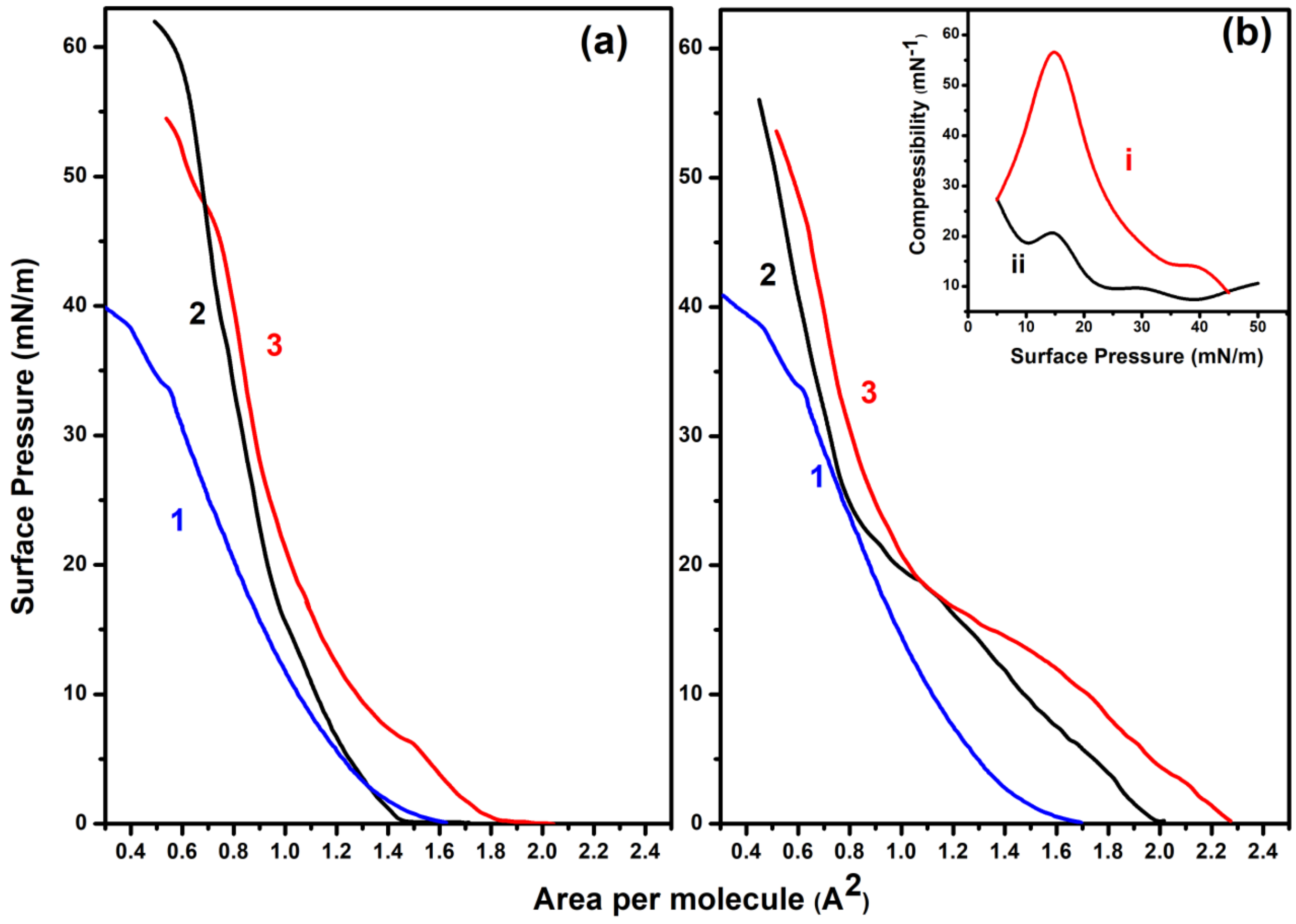

**Fig. 1.**

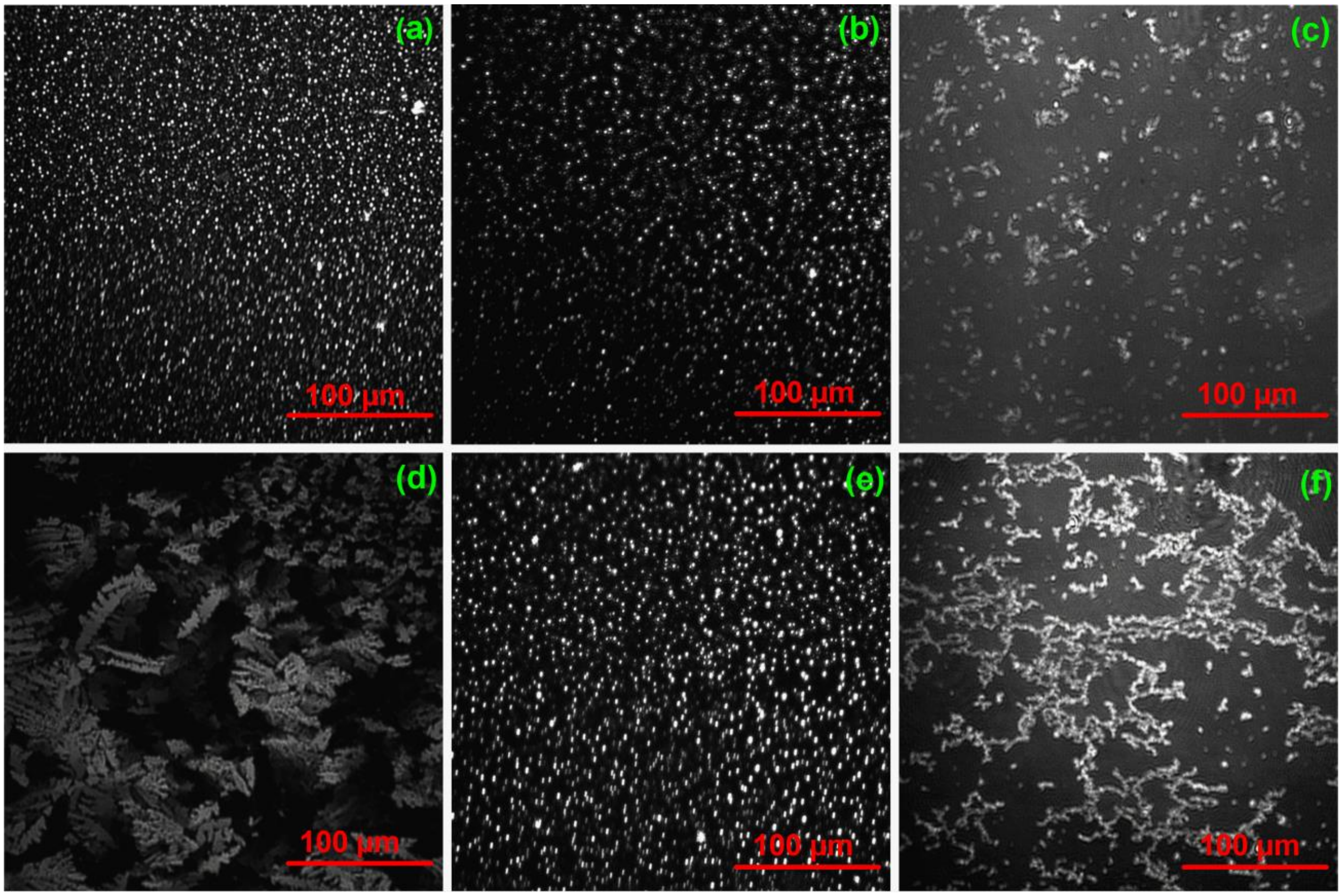

**Fig. 2.**

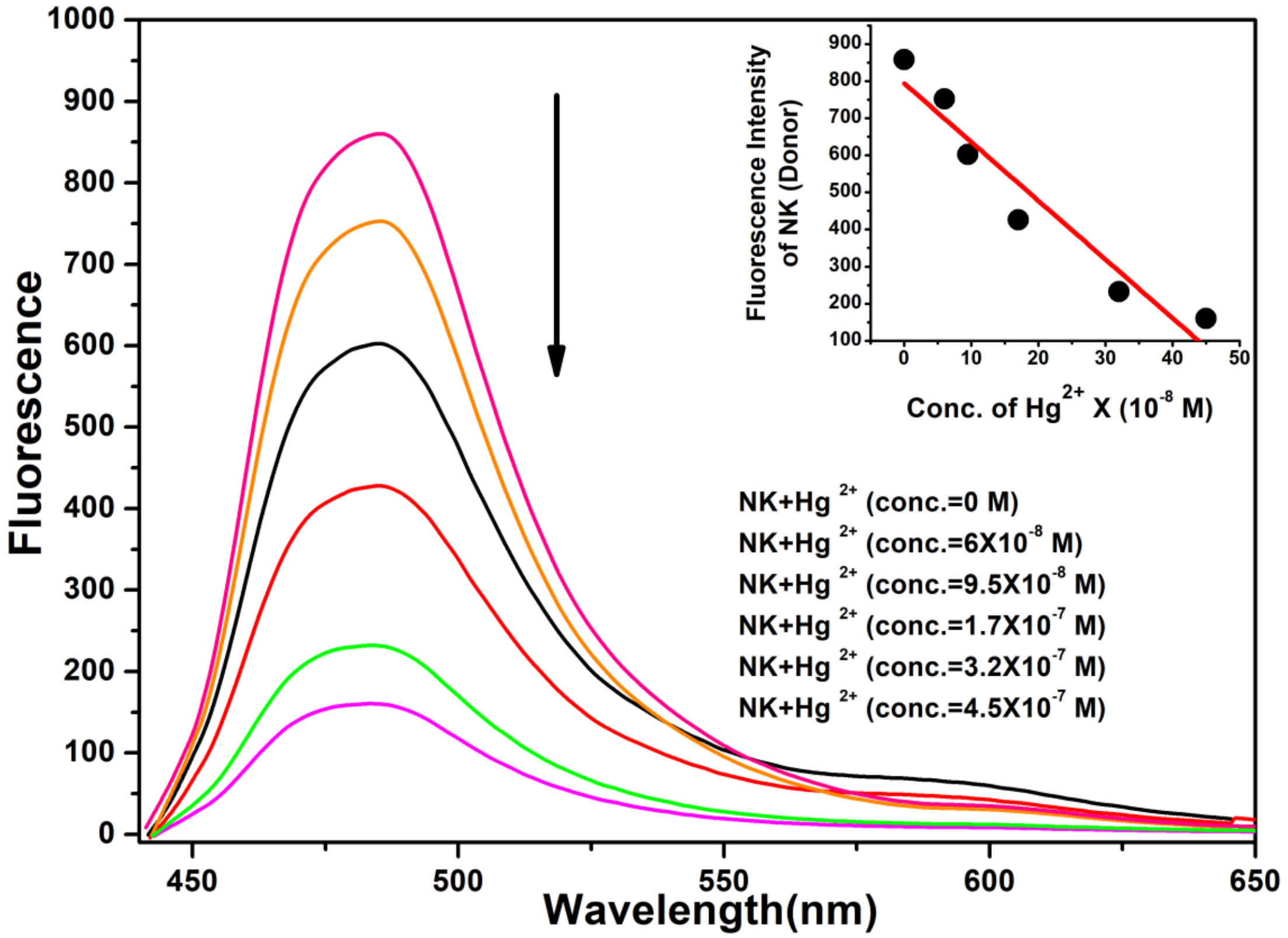
Fluorescence
Wavelength(nm)
1000
900
800
700
600
500
400
300
200
100
0
450
500
550
600
650
Fluorescence Intensity of NK (Donor)
Conc. of Hg2+ X (10-8 M)
NK+Hg 2+ (conc.=0 M)
NK+Hg 2+ (conc.=6X10-8 M)
NK+Hg 2+ (conc.=9.5X10-8 M)
NK+Hg 2+ (conc.=1.7X10-7 M)
NK+Hg 2+ (conc.=3.2X10-7 M)
NK+Hg 2+ (conc.=4.5X10-7 M)

**Fig. 3a.**

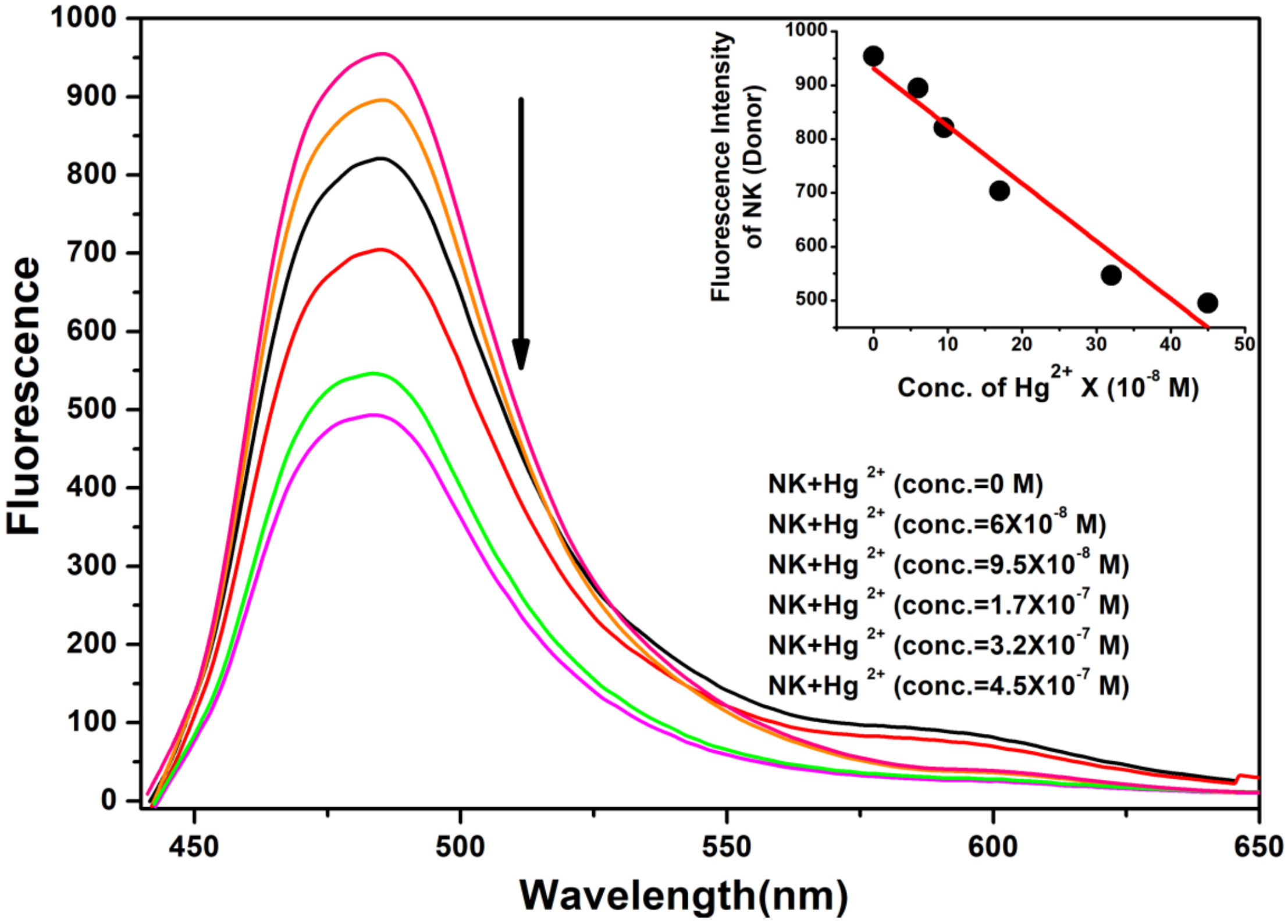

Fluorescence
Wavelength(nm)
Fluorescence Intensity of NK (Donor)
Conc. of Hg 2+ X (10-8 M)
NK+Hg 2+ (conc.=0 M)
NK+Hg 2+ (conc.=6X10-8 M)
NK+Hg 2+ (conc.=9.5X10-8 M)
NK+Hg 2+ (conc.=1.7X10-7 M)
NK+Hg 2+ (conc.=3.2X10-7 M)
NK+Hg 2+ (conc.=4.5X10-7 M)

**Fig. 3b.**

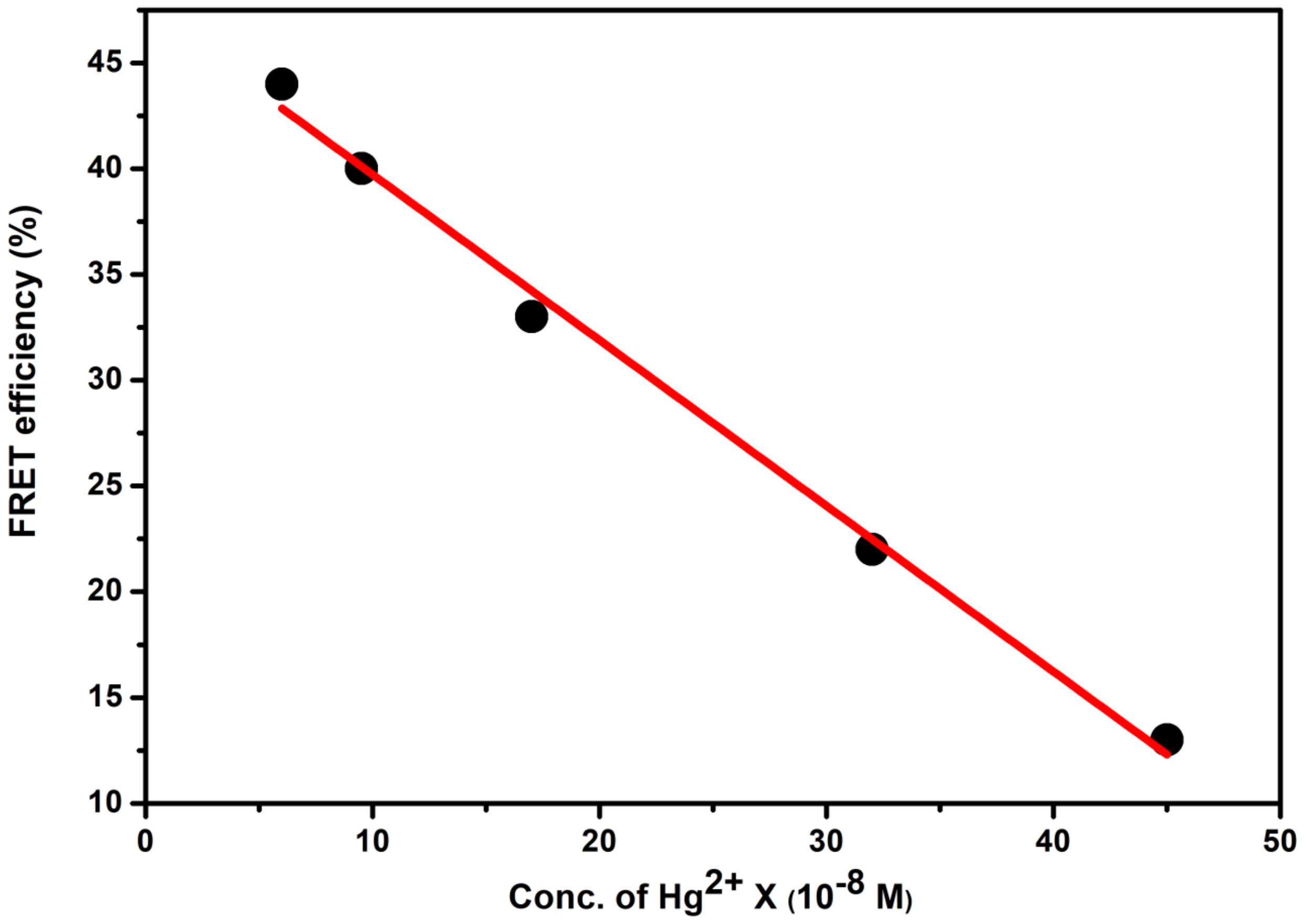

**Fig. 4a.**

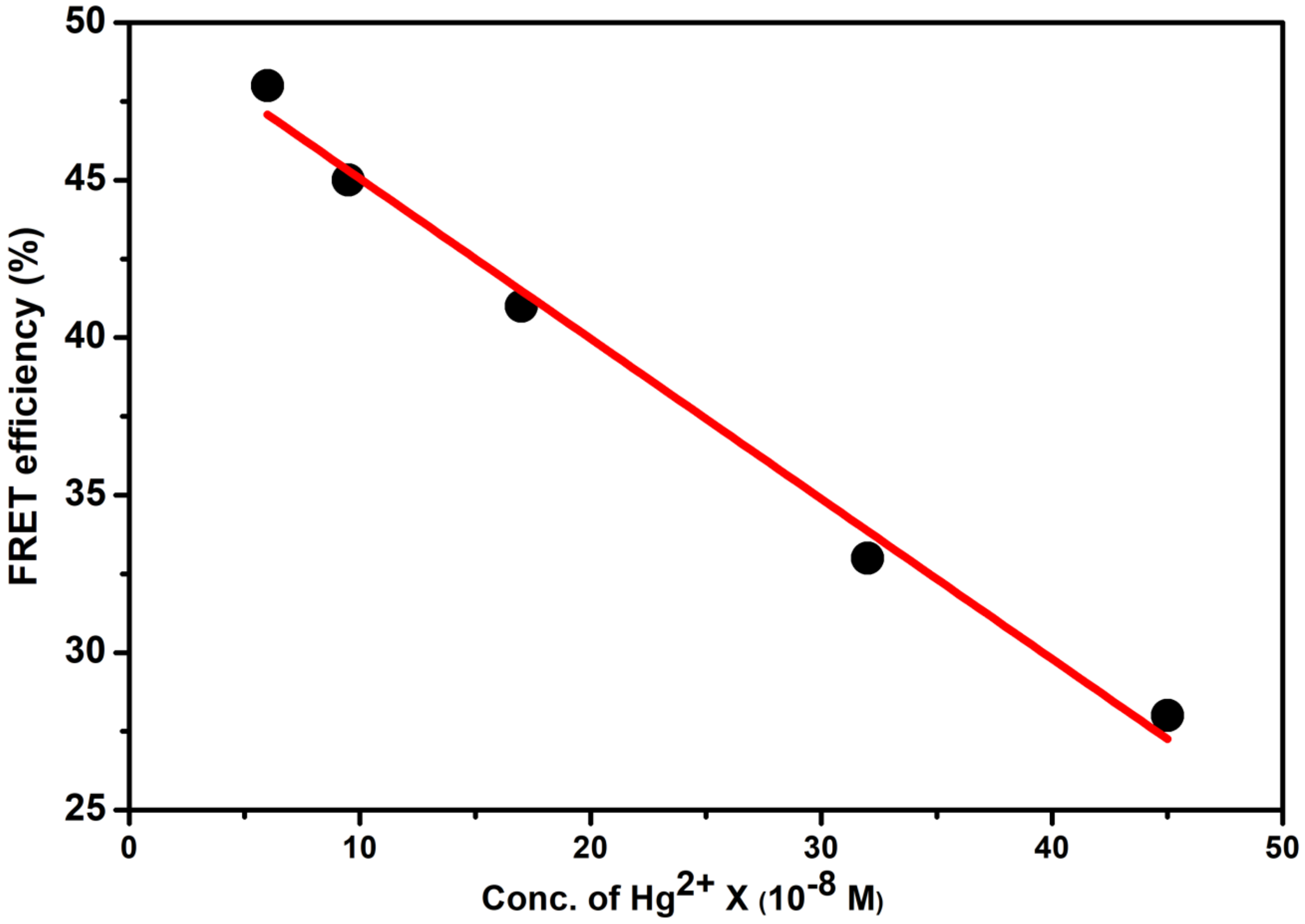

**Fig. 4b.**

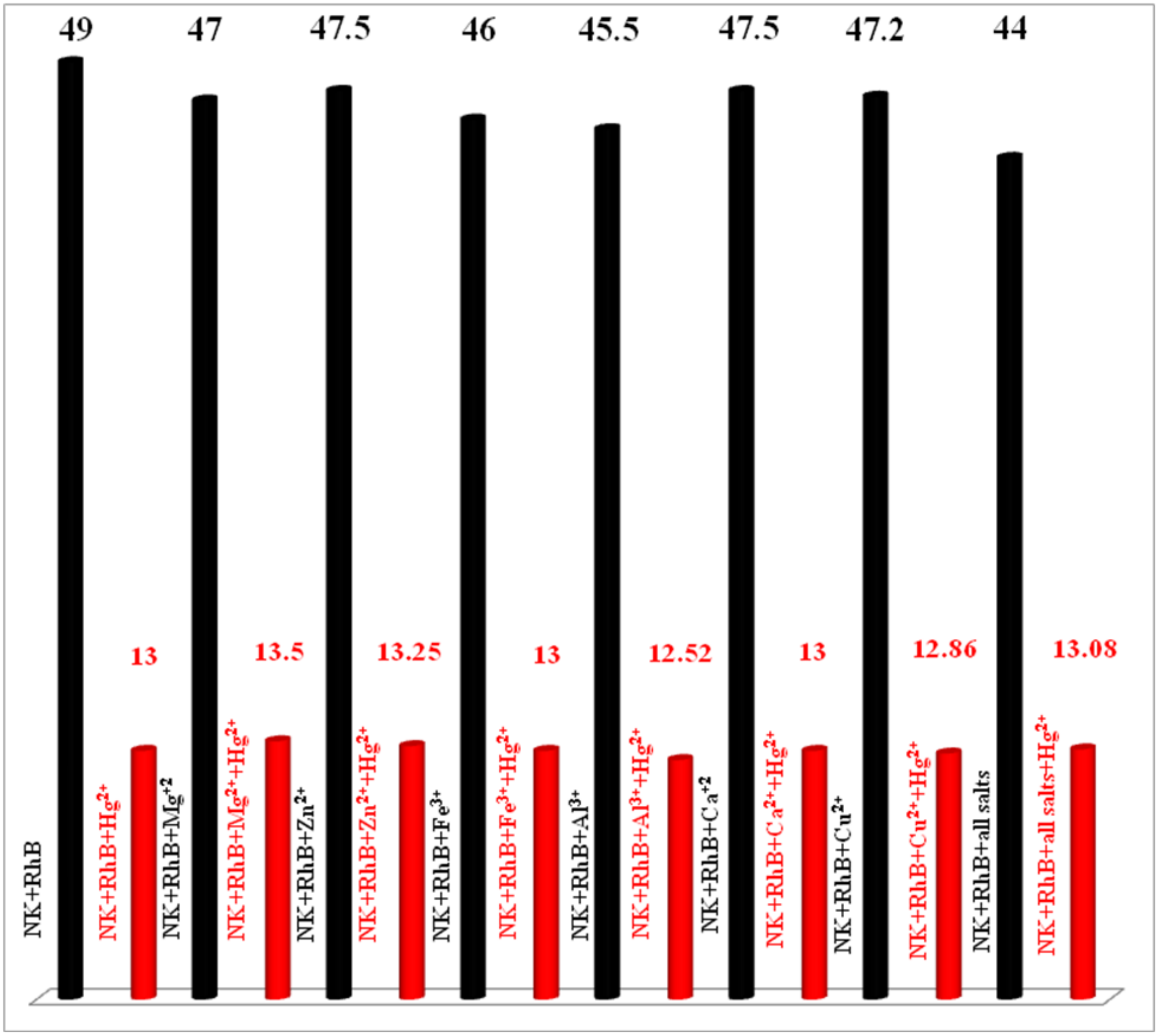

**Fig. 5.**

**Table caption**

**Table 1.** Spectroscopic characteristics of NK and RhB in LB and SC films.

**Table 2.** Values of spectral overlap integral (J(λ)), Förster radius ($R_0$), donor–acceptor distance (r) and energy transfer efficiency (E%) for FRET between NK and RhB with different $Hg^{2+}$ concentration in LB films. Concentrations of both the dyes are $0.5\times10^{-3}$ M. Excitation wavelength is 430 nm. These are calculated from the spectral characteristics of Fig. S3 and S5a.

**Table 3.** Values of spectral overlap integral (J(λ)), Förster radius ($R_0$), donor–acceptor distance (r) and energy transfer efficiency (E%) for FRET between NK and RhB with different $Hg^{2+}$ concentration in SC films. Concentrations of both the dyes are $0.5\times10^{-3}$ M. Excitation wavelength is 430 nm in SC films. These are calculated from the spectral characteristics of Fig. S3 and S5b.

**Table 4.** LOD values in SC and LB films for the sensors based on fluorescence as well as FRET technique calculated from Fig. 3 and 4.

**Table 5.** Variation of FRET efficiency in presence and absence of salt and $Hg^{2+}$. Concentrations of NK and RhB are fixed at $0.5\times10^{-3}$ M. Concentration of $Hg^{2+}$ and different salts are $4.5\times10^{-7}$ M.

**Table 6.** Determination of $Hg^{2+}$ in sample waters.

**Tables**

**Table 1**

| | Absorption | | Fluorescence | |
|---|---|---|---|---|
| | NK | RhB | NK | RhB |
| **SC film** | 423, 442 | 539, 580 | 484 | 596 |
| **LB film** | 410, 433, 461 | 532, 573 | 484 | 600 |

**Table 2**

| **$Hg^{2+}$ concentration (in M)** | **$J(\lambda)\times10^{15}$ $M^{-1}cm^{-1}nm^4$** | **$R_0$ (nm)** | **r (nm)** | **FRET efficiency (%)** |
|---|---|---|---|---|
| 0 | 4.48 | 6.118 | 6.158 | 49 |
| $6\times10^{-8}$ | 4.09 | 6.026 | 6.273 | 44 |
| $9.5\times10^{-8}$ | 3.75 | 5.939 | 6.354 | 40 |
| $1.7\times10^{-7}$ | 3.12 | 5.760 | 6.481 | 33 |
| $3.2\times10^{-7}$ | 2.32 | 5.483 | 6.770 | 22 |
| $4.5\times10^{-7}$ | 2.09 | 5.388 | 7.396 | 13 |

**Table 3**

| **$Hg^{2+}$ concentration (in M)** | **$J(\lambda)\times10^{15}$ $M^{-1}cm^{-1}nm^4$** | **$R_0$ (nm)** | **r (nm)** | **FRET efficiency (%)** |
|---|---|---|---|---|
| 0 | 4.80 | 6.189 | 6.066 | 53 |
| $6\times10^{-8}$ | 4.50 | 6.123 | 6.205 | 48 |
| $9.5\times10^{-8}$ | 4.12 | 6.033 | 6.238 | 45 |
| $1.7\times10^{-7}$ | 3.99 | 6.001 | 6.376 | 41 |
| $3.2\times10^{-7}$ | 3.54 | 5.883 | 6.620 | 33 |
| $4.5\times10^{-7}$ | 3.21 | 5.788 | 6.774 | 28 |

**Table 4**

| | **LB** | | **SC** | |
|---|---|---|---|---|
| | **Adj. $R^2$** | **LOD (ppb)** | **Adj. $R^2$** | **LOD** |
| **Fluorescence** | 0.886 | 37.64 | 0.927 | 29.53 |
| **FRET** | 0.992 | 09.13 | 0.988 | 11.70 |

**Table 5**

| Sample | Efficiency (%) |
|---|---|
| (NK + RhB) + $Mg^{2+}$ | 47.00 |
| NK + RhB + $Mg^{2+}$ + $Hg^{2+}$ | 13.50 |
| (NK + RhB) + $Zn^{2+}$ | 47.50 |
| NK + RhB + $Zn^{2+}$+ $Hg^{2+}$ | 13.25 |
| (NK + RhB) + $Fe^{+3}$ | 46.00 |
| NK + RhB + $Fe^{+3}$+ $Hg^{2+}$ | 13.00 |
| (NK + RhB) + $Al^{+3}$ | 45.50 |
| NK + RhB + $Al^{+3}$+ $Hg^{2+}$ | 12.52 |
| (NK + RhB) + $Ca^{2+}$ | 47.50 |
| NK + RhB + $Ca^{2+}$+ $Hg^{2+}$ | 13.00 |
| (NK + RhB) + $Cu^{2+}$ | 47.20 |
| NK + RhB + $Cu^{2+}$+ $Hg^{2+}$ | 12.86 |
| NK + RhB + water | 49.00 |
| NK + RhB + $Hg^{2+}$ | 13.00 |
| (NK +RhB) + (All salts) | 44.00 |
| (NK +RhB) + $Hg^{2+}$+(All salts) | 13.08 |

**Table 6**

| Sample | Added ($\times 10^{-7}$ M) | Found ($\times 10^{-7}$ M) | Recovery (%) | RSD(%, n=3) |
|---|---|---|---|---|
| 1# | 1.70 | 1.65 | 97.05 | 2.2 |
| | 3.20 | 3.10 | 96.80 | 0.6 |
| | 4.50 | 4.45 | 98.80 | 1.3 |
| 2# | 1.70 | 1.67 | 98.23 | 0.9 |
| | 3.20 | 3.09 | 96.56 | 1.2 |
| | 4.50 | 4.32 | 96.00 | 1.9 |
| 3# | 1.70 | 1.62 | 95.29 | 0.7 |
| | 3.20 | 3.10 | 96.87 | 2.1 |
| | 4.50 | 4.37 | 97.11 | 1.5 |